# Minimum Capital Requirement Calculations for UK Futures

**JOHN COTTER***
University College Dublin

**Address for Correspondence:**
Dr. John Cotter,
Director of the Centre for Financial Markets,
University College Dublin,
Blackrock,
Co. Dublin,
Ireland.

E-mail. john.cotter@ucd.ie
Ph. +353 1 706 8900
Fax. +353 1 283 5482

* The author would like to thank Colm Kearney, Donal McKillop, Rool Oomen, Enrique Sentana, Neil Shephard, the referee, and participants at the Royal Economic Society's Econometrics seminar at Nuffield College, University of Oxford for their comments. The research received financial support from CPA Ireland.

# Minimum Capital Requirement Calculations for UK Futures

**Abstract**

Key to the imposition of appropriate minimum capital requirements on a daily basis requires accurate volatility estimation. Here, measures are presented based on discrete estimation of aggregated high frequency UK futures realisations underpinned by a continuous time framework. Squared and absolute returns are incorporated into the measurement process so as to rely on the quadratic variation of a diffusion process and be robust in the presence of fat tails. The realized volatility estimates incorporate the long memory property. The dynamics of the volatility variable are adequately captured. Resulting rescaled returns are applied to minimum capital requirement calculations.

# Minimum Capital Requirement Calculations for UK Futures

## I. Introduction:

This paper offers volatility measures for UK futures traded on the London International Financial Futures and Options Exchange (LIFFE) underpinned by the continuous time theoretical finance literature (for a discussion of innovations in this area see Andersen et al, 2001). These measures are used to rescale returns for minimum capital requirement calculations. The volatility measures help to bridge the gap between the continuous time stochastic differential equations systems that play such an important role in asset pricing models such as Black-Scholes, and the discrete time stochastic difference equation models such as GARCH related processes' popularly employed in empirical finance. Of greater importance to this study is that this same gap exists between the stochastic differential equations systems and the discrete approximation models used in risk management estimations such as Value at Risk (VaR) type estimates (Neftci, 2000). This gap results from the conditional variances and covariances being inherently unobservable and the requirement of VaR measures to provide accurate estimates of these variables.

A secondary related problem-facing regulators results from how they usually try to address the unknown conditional variance issue. Usually, this involves assuming a particular distribution, for example normality, for the inputs or returns that are then used to give ex-post measures for the unknown volatility in a VaR framework. However, financial futures returns generally have fat tails and do not correspond to these assumed distributions resulting in estimation problems (Taylor, 1986). This paper addresses this issue by obtaining rescaled returns that are approximately gaussian allowing us to deliver conservative and accurate risk measures. In treating the fat-tailed property of futures returns, it is appropriate to match this empirical feature in the estimation of conditional volatility. Davidian and Carroll (1987) find that absolute realizations are more robust to the presence of fat-tailed observations than their squared counterparts, and this finding is implemented by extending the theoretical framework outlined for squared realizations, to absolute ones.[1]

This paper provides model-free volatility estimates building on the quadratic variation of a diffusion process for aggregated squared and absolute realizations. The diffusion process allows for accurate realized volatility estimation, as the sample interval becomes negligible. Furthermore, the choice of realizations demonstrates the empirical stylized facts of financial time series, with fat-tailed distributions and volatility clustering. An illustration of the methodology is applied to risk management estimates for different asset types with varying degrees of risk. Specifically, minimum capital requirements are calculated at various probability levels for the futures with short and long trading positions. Minimum capital requirements are part of a statutory regulating measure should be seen separately from the margin requirements imposed by the exchange on which the contracts are traded (see Cotter (2001) for an illustration).

Minimum capital requirements are set by regulatory bodies to cover market risk of financial firms that protects against losses arising from the volatility of their holdings.[2] Regulators impose minimum capital requirements to avoid systemic costs of default by reducing the probability of failure. There are a number of alternative models in place for calculating minimum capital requirements. These include the Comprehensive approach of the, amongst others, US Securities and Exchange Commission, and of particular relevance to the assets analyzed in this paper, the Portfolio approach of the UK's Securities and Futures Authority. The approaches are similar. The Block Building approach incorporates two aspects in measuring minimum capital requirements. It primarily focuses on a calculation on the net trading position, recognizing the risk reduction caused by offsetting long and short investments. In addition, it adjusts the net measure upwards by an arbitrary percentage assuming that diversification does not eliminate the unique risk of investments. The Portfolio approach also incorporates both elements of risk reduction, but relies only on portfolio theory where capital requirements reflect the risk facing investors through statistical measures of volatility, for example the unconditional standard deviation, in a Value at Risk (VaR) type mechanism.[3]

Imposing minimum capital requirements and at what level, involves a trade-off between the costs incurred by financial firms and their effect on competition in the market place, and the costs borne by society due to non-fulfillment of contracts by the security firms. Taking the first part of this compromise, large capital requirements have to be borne by some economic agent whether it is the security firm or investors. Also, increased costs may affect the competitive nature of the securities markets by imposing entry barriers to the industry, ultimately affecting investors' transaction costs. Turning to the social costs resulting from default risk, there are many costs of negative externalities occurring. Looking at the asset class chosen in this paper, the disasters for derivative firms such as Barings, Daiwa, Metallgesellschaft, and Orange County provides an incentive to try to avoid these losses.[4] In reality, regulators have usually agreed on taking a practical course of action that sets capital requirements at such a level that default risk is controlled for, but not eliminated.[5]

Notwithstanding the approaches in place, minimum capital requirement methodologies should focus on two key issues, the volatility inherent in the investment position, and the holding period involved. Accurate volatility estimation for derivative positions is essential to imposing optimal capital deposits, and this paper concentrates on this topic. It provides alternative volatility measures and discusses their empirical features, and then proceeds to calculate minimum capital requirements for three futures contracts traded on LIFFE. This methodology follows the Portfolio approach by incorporating a VaR type model and estimating the potential effects of futures realized volatility at different probability levels. Essentially, the paper is trying to establish the probability of default with associated minimum capital requirement estimates.

The outline of the paper proceeds as follows. In the next section a brief description of the theoretical framework underpinning the continuous time realized volatility measures is outlined. In addition, the actual process for obtaining the daily aggregate measures is given. Section III describes the assets analyzed and their data capture. Previously, there are studies available on stock index and bond contracts using this measurement approach. However, this is the first study to analyze short run interest rate series'. Section IV presents the empirical features of the daily return and volatility measures detailing properties from their unconditional and conditional distributions. This is followed in section V with a risk management application describing the minimum capital requirements for both long and short positions in the futures at various confidence intervals. By way of contrast, gaussian estimates are also shown. Finally a summary and conclusions for the paper are given in section VI.

**II. Theoretical Framework:**

*A. Continuous Time Volatility Measurement:*

Much of the theoretical framework is described in Anderson et al (2001) and their related papers. This section provides a brief synopsis of the approach plus any deviations and extensions applicable to this study. First take integrated volatility as the measure of volatility in a continuous time process. This variable is unobservable, but it is shown that it is equivalent to realized volatility, which is observable. Assuming a continuous time process which generates instantaneous returns, $r_t = dp_t$ where $p_t$ is the logarithmic price process, and the diffusion process is given as:

$$(1) \qquad dp_t = \mu(p_t, \sigma_t) + \sigma_t dw_t$$

for $w_t$ is brownian motion, $t \geq 0$, and the functions $\mu(p_t, \sigma_t)$ and $\sigma_t$ are strictly positive. The drift term, $\mu(p_t, \sigma_t)$, may hereafter be excluded from the continuous time process as empirically the conditional means is found to be close to zero, and with little dependence for high frequency realizations. The approach assumes that the return process is observable, whereas in contrast the volatility process is not. The diffusion process in (1) allows us to obtain greater accuracy in the estimation of the conditional variance if the diffusion is observed at finer intervals. At the limit, when interval size goes towards zero, the observed diffusion allows exact estimation of the conditional variance.[6] Thus regulators now have volatility estimates that can be treated as error free.

Discrete approximations of the process in (1) using high frequency data have $r_{m,t} = p_t - p_{t-1/m}$ as the continuously compounded returns with m evenly spaced observations per day. The unit interval represents one day with 1/m horizon returns.[7] Alternative strategies are available for the actual estimation of realized volatility including block sampling and rolling regression. The former relies on splicing the realizations into even sized block and treating realized volatility as constant within a block (see Merton (1980) using interdaily returns; and Schwert (1990a) using intraday realizations). The latter applies rolling sample estimates to the returns series' (for high frequency observations see Foster and Nelson (1996); and Campbell et al (2001) for lower frequency realizations). Block samples of m observations are taken at 5-minute intervals aggregating out to a daily measure used later to provide accurate daily risk management estimates. The choice of 5-minute intervals is done on the basis of a trade-off between microstructure arguments and obtaining outcomes from a diffusion process (see Goodhart and O' Hara, 1997; for a general discussion of microstructure issues). These imply that by obtaining realizations of the highest possible frequency so that if m → ∞, continuously measured returns are obtained.[8] Assuming $\sigma_t$ and $w_t$ are independent, the variance for an horizon H of the stochastic process of $\sigma_t$ with cadlag sample paths (continuous from the right with limits from the left) is given by

(2) $$\sigma^2_{t,H} \equiv \int_0^H \sigma^2_\tau d\tau$$

This implies that $\sigma_t^2$ has continuous sample paths. The quadratic variation of (2) is a semi-martingale, although other approaches could also be used, for example, the non-gaussian Ornstein-Uhlenbeck processes that also allows for a characterization of their quadratic variation (see Barndorff-Nielsen and Shephard (2001)). Also, this integrated volatility process is equated to the quadratic variation of returns over a time interval from t to t + H:

$$(3) \quad \plim_{m \to \infty} \left\{ \int_{t-H}^{t} \sigma_\tau^2 d\tau - \sum_{j=1,\ldots,m} r_{m,t+j/m}^2 \right\} = 0$$

Implying for m sampling frequency, the realized volatility is consistent for the integrated volatility. Assuming that the sample returns are white noise and $\sigma^2_t$ has continuous sample paths allows the theory of quadratic variation to imply that the limiting difference between the unobserved volatility estimate and the observed realizations of the returns process be zero (Karatzas and Shreve (1991)).

Moving to another class of integrated volatility estimators uses aggregated absolute returns, $\sum |r_m|$ and its' variants. Switching the attention to the theoretical framework, the integrated process, $\int \sigma^2$, is now in terms of aggregated absolute realizations:

$$(4) \quad s_{t,H}^2 \equiv \int_0^H |r_t| dt$$

Assuming that they measure the time variation in the diffusion process of (1). These absolute return estimates are particularly relevant in the context of the commonly cited fat-tailed behavior as they examine the absolute extent of price variation (see for example, Davidian and Carroll (1987) and Granger and Ding (1995)). These studies indicate that absolute realizations model fat-tailed characteristics relatively better than their squared counterparts. In contrast in the use of thin-tailed distributions, squared realizations may perform substantially better than absolute ones as they are more robust to thin-tailed distributions.

*B. Aggregated Daily Measures:*

Turning now to the practical implementation of aggregate measures for any day t, let an intraday interval be measured by m. The number of intraday intervals will be asset dependent varying according to the associated trading hours. The assets are traded on a 'working day' cycle whereas in contrast, currencies are traded on a 24-hour cycle. Each day's returns, $r_t$, are obtained by aggregating the high frequency intraday returns, $r_{m,t}$:

$$(5) \qquad r_t = \sum_{j=1}^{m} r_{m,t+j/m}$$

Similarly, the daily volatility measures are obtained by aggregating variations of the intraday realizations. The chosen measures, the squared returns (similar to an asset's variance) - $[r_t^2]$; their square root (similar to an asset's standard deviation) - $[\sqrt{r_t^2}]$; the absolute returns - $|r_t|$; and the absolute returns raised to the power coefficient of one half - $|r_t|^{0.5}$; are the realizations of the daily realized volatility series'. For example the first two realized volatility measures using squared realizations are given as:

$$(6) \qquad [r_t^2] = \sum_{j=1}^{m} [r_{m,t+j/m}^2]$$

and

$$(7) \qquad [\sqrt{r_t^2}] = \sum_{j=1}^{m} [\sqrt{r_{m,t+j/m}^2}]$$

Turning now to the extension for fat-tailed characteristics, two further daily realized volatility measures using absolute realisations are presented:

$$(8) \qquad |r_t| = \sum_{j=1}^{m} |r_{m,t+j/m}|$$

and

$$(9) \quad |r_t^{0.5}| = \sum_{j=1}^{m} |r_{m,t+j/m}^{0.5}|$$

A weighting scheme may be implemented for each volatility measure to emphasise some impact of intraday realisations, but here it is assumed that the weights are equal within a daily trading block.

### III. Data Features:

Daily measures are obtained using the aggregation method described in the previous section. The assets chosen, the FTSE100, the UK Long Gilt and Sterling contracts are the most traded representatives of the Stock Index, Bond and Interest Rate futures on the LIFFE exchange. As the three futures represent different asset types, their calculations are based on diverse criteria. The FTSE100 contract is based on calculations of the asset in the underlying spot market, the UK Long Gilt contract is a future on a notional 10 year UK government bond with a 7% coupon, and the Sterling interest rate contract is based on the three month London Inter-Bank Offer Rate (LIBOR), the wholesale rate on which financial institutions borrow and lend from each other. The theoretical underpinnings have not previously been applied to this latter type asset in the literature. Benchmark studies are however available for the former asset types (see for example, Andersen et al (2000a), (2000c) and Areal and Taylor (2002) for the stock market; and Bollerslev et al, (2000) for the bond market).

The period of analysis is from January 1 1996 through to December 31 1999 leaving a sample of 1003 full trading days. A number of issues arise in the data capture process. First, all holidays are removed. This entails New Year's (two days), Easter (two days), May Day, Spring holiday (1 day), Summer holiday (1 day), and Christmas (two days).[9] In addition, trading took place over a half day during the days prior to the New Years and Christmas holidays and these full day periods are removed from the analysis. 5-minute intervals are chosen for analysis. This is based on a trade-off between microstructure arguments and trying to obtain continuous realizations as m $\to \infty$.[10] It is worth noting some of the possible microstructure features affecting the empirical implementation of the theoretical framework described in the last section. In addition to bid-ask spreads index returns feature positive serial correlation due to non-synchronous trading effects.[11] Moreover, volatility levels vary from trading to non-trading periods, and volatility patterns change in an intraday basis. This paper follows the analysis completed in a series of papers and chooses 5-minute intervals (for example, Andersen et al (2001) and Andersen et al (2000c)). Although the assets chosen for analysis are the most actively traded UK futures, the level of trading activity may become an issue for other thinly traded assets. Measurement error and the level of liquidity are negatively related so caution should be followed in the analysis of less active traded contracts.

Each contract has different trading hours resulting in a unique number of daily 5-minute intervals – FTSE100 (113), UK Long Gilt (120), and Sterling (118). With 1,003 trading days, the analysis covers 113,339 intervals for the stock index, 120,360 intervals for the bond, and 118,354 intervals for the interest rate futures' assuming a 5-minute returns series, $[r_{m, t}]$, involves t = 1, 2,…., 1003 days, and m, the number of respective contract's daily intervals. Also, each contract has four expiration periods, March, June, September, and December with data being chosen for the period nearest to delivery.[12] The returns series', [$r_{m,t}$], are calculated using the first difference of each interval's log closing price, as in $[\ln(P_{m,t}) - \ln(P_{m-1,t})]$. The daily returns series, [$r_t$], are obtained by summing m intraday interval realizations for each asset. However, intraday periods may be liable to thin trading, and if a trade does not occur for any 5-minute interval, the interpolated bid/ask spread from that period is used if available, or otherwise the previous period's value is used.[13] This results in the possibility of consecutive returns remaining equivalent in magnitude.

**IV. Empirical Features of Returns and Volatility Measures:**

*A. Unconditional Distribution of Daily Aggregate Measures:*

The unconditional daily returns series, [$r_t$], using the aggregation method is examined for their properties. Summary statistics and box plots are given for each futures contract in table I. On average, returns are positive for the stock index and bond contracts over the period of analysis with the FTSE100 offering the highest return. The respective sample means for each asset are 0.071% (FTSE100), 0.010% (UK Long Gilt) and $-1.424*10^{-4}$% (Sterling) corresponding to annual averages with 251 trading days of 17.821% (FTSE100), 2.510% (UK Long Gilt) and 0.036% (Sterling). The percentiles represent possible realizations on short and long positions with the lower percentiles giving losses on long positions whereas the upper values give losses on short ones. The importance of the percentiles will become apparent when an illustration of the aggregated return and volatility measures in a risk management context is given. Here the minimum capital requirement needed to cover the expected losses being expressed for a variety of confidence levels using percentiles will be shown. Moving onto the scale measures, generally the statistics emphasize a large degree of deviation for the realizations of different asset types. The two unconditional volatility measures, the standard deviation and variance also diverge according to asset type with the interest rate contract being least risky and the stock index contract being most risky. All contracts exhibit excess skewness and kurtosis with the short run interest rate contract behaving least well. For example, the leptokurtotic returns involves a bunching of realizations around the median, and more importantly for regulators trying to minimize default risk, a large number of values clustering around the tails of the distribution. The paper will determine later if the returns series' can be transformed to have more attractive time series properties for the development of accurate risk management estimates, namely a symmetric iid variable.

INSERT TABLE I HERE

Many related volatility proxies can be calculated with the method. Andersen et al (2000c) find log realizations of aggregated squared returns almost gaussian. Thus, as well as the four variations of absolute and squared returns, log realizations of these measures are also included. The assumption of zero first moments is commonly applied in this type of analysis, although it is easy to scale the realized volatility estimates by a mean return measure. Evidence in support of applying this assumption is detailed in table I. Given an assumption of zero first moments, the mean of $[r_t^2]$ is closely associated with the commonly used variance estimate.[14]

By extension, the other commonly cited risk measure, the standard deviation, is represented by $[\sqrt{r_t^2}]$. Both absolute realizations indicate dispersion and take account of the size of the fluctuations in returns regardless of whether they are positive or negative. Empirically, the magnitude of financial return fluctuations often are very large vis-à-vis a normal distribution giving support for a fat-tailed property for these realizations. In table II, some summary statistics of the volatility measures using the aggregated returns method are presented. As can be seen, the magnitude of the realized volatility values is large for the relatively risky stock index futures. For example, the average daily value using $|r_t|$ is 7.339% and for the more commonly cited measure of dispersion, the standard deviation is 1.071%. The interest rate contract indicates the lowest realized volatility levels. Absolute return volatility measures dwarf their squared returns counterparts.

INSERT TABLE II HERE

Two dispersion measures of the newly formed realized volatility series, the variance and standard deviation, suggest that the actual divergence in the respective futures second moment properties is generally small, giving greater statistical inference to risk findings. For example, the $|r_t|$ measure is relatively noise free (0.049% as given by the variance) for the UK Long Gilt contract given its average risk levels (2.994%). In terms of distributional shape, the unconditional $3^{rd}$ moment estimates clearly suggest extremely right skewed variables in all cases. The conclusions regarding the $4^{th}$ unconditional moment is not as clear-cut. In general, while financial returns exhibit a fat tail property with excess kurtosis, this does not occur for the UK Long Gilt contract. However, all exhibit a non-gaussian clustering of realisations. Andersen et al (2000b) find that the lack of support for a non-gaussian dependency structure in the daily volatility measures is a result of the strong dependence in the intraday values. Here, all series are leptokurtotic with the exception of the $|r_t|^{0.5}$ series for the UK Long Gilt which is platykurtotic. Log realisations of the four realized volatility measures are also included in table II to determine their shape. Generally, the excess kurtosis is reduced for these series. Figure 1 shows the box plots for the series with optimal shape properties of all realisations for each series.[15] These are respectively the $Ln|r_t|$ series for the FTSE100 and UK Long Gilt contracts and the $Ln|r_t|^{0.5}$ series for Sterling. In the two former series, the fat-tail characteristic is removed, although excess skewness is still exhibited. Nonetheless, gausssian iid behaviour for these volatility measures is almost achieved.[16]

INSERT FIGURE 1 HERE

B. Conditional Distribution of Aggregate Measures:

Whilst similar claims can be made for each asset's conditional returns series there are some variations in terms of conditional findings for the assets when examining their volatility patterns.[17] Also, there are distinctions according to the different realized volatility measures. Beginning with the common features, volatility clustering is shown in the time series plots of figure 2 where the series' $|r_t|$ are used for illustration. The plots indicate clearly that volatility varies over time and that the magnitudes of the realisations tend to bunch together with periods of relatively low and high values.

INSERT FIGURE 2 HERE

Turning next to the memory characteristics, empirical analysis of financial time series suggests that the long memory feature dominate for absolute over squared realisations (see Ding and Granger (1996)). Thus, long memory is investigated for the daily realized volatility series, $|r_t|$, by calculating the degree of fractional integration, d, for $0 \leq d \leq 0.5$. The characteristic implies that the rate of decay for any lag, $\tau$, is given by $\tau^{2d-1}$. The long memory estimate is measured using the Geweke and Porter-Hudak (1983) log-periodogram regression approach, $d_{GPH}$, updated for non-gaussian volatility estimates by Deo and Hurvich (2000). This adjustment is required given the fat-tailed and right skewed behaviour of the volatility series. Assuming, $I(\omega_j)$ stands for the sample periodogram at the $j^{th}$ fourier frequency, $\omega_j = 2\pi j/T$, j = 1, 2, ..., [T/2], the log-periodogram estimator of $d_{GPH}$ is based on regressing the logarithm of the periodogram estimate of the spectral density against the logarithm of $\omega$ over a range of frequencies $\omega$:

(10) $$\log[I(w_j)] = b_0 + b_1 \log(w_j) + U_j$$

where j = 1, 2,..., m, and $d = -1/2\beta_1$. Deo and Hurvich (2000) note a trade-off between the variance and bias of the least squares regression estimator, where with increasing m, the bias increases and the variance decreases. Also, estimates of d are dependent on the choice of m. Taking these two issues into account, estimates of $d_{GPH}$ are obtained by using $m = T^{4/5}$ as suggested by Andersen et al (2001). This implies that for $m = T^{\delta}$, a sample of 252 periodogram estimates is employed. In addition, standard errors are reported but may be problematic as an asymptotic normal limiting distribution exists for the long memory estimator if and only if $\delta < (1 + 4d)^{-1} 4d$ which can change depending on the value of d (Deo and Hurvich (2000)).[18] The asymptotic standard errors are given by $\pi(24m)^{-1/2}$, that is 0.040.

INSERT FIGURE 3 HERE

First, long memory is demonstrated in figure 3 where the strong persistence that follows a hyperbolic decay structure is shown. Also, all the corresponding Ljung-Box statistics are significant given in table III. It is important to note that this long memory feature occurs for a relatively small data set and for a small number of lags (in contrast for example, Granger and Ding (1995) use Schwert's (1990b) data set of 17055 observations) and shows an advantage of the aggregation of intraday observations. There is a cyclical pattern with the overall decay structure of realized volatility in figure 3. This is due to day of the week and intraday effects. These effects are demonstrated by taking a 5-day representation of the lag structure in figure 4. Interdaily, a reduction in dependence moving through a week with a slight increase at the end of the cycle is exhibited. Also, overall the intraday pattern is u-shaped with a self-similar pattern during the course of the day. These correspond in general to the findings for equity and fixed income markets (see Andersen et al (2000a) and Bollerslev et al (2000)). Causes for these interday and intraday stylized facts include effects from macroeconomic news announcements on varying volatility and autocorrelation patterns for intraday realisations, and the impact on volatility magnitude in terms of non-trading days on the subsequent day's estimate.

INSERT FIGURE 4 HERE

INSERT TABLE III HERE

The long memory estimates, $d_{GPH}$, are given in table III supporting a fractionally integrated process. The values for the stock index and bond contracts are similar in magnitude to previous evidence for comparable assets where short memory processes are also rejected. For example, Bollerslev et al (2000) find log periodogram estimates of between 0.35 and 0.45 for US treasury bond and equity assets. Andersen et al (2001) give an implication of this result and note that short memory modeling, for example, assuming that volatility follows an Ornstein-Uhlenbeck process, which is commonly applied in the theoretical finance literature, is incorrect. In contrast, long memory modeling should be applied. However, the Ornstein-Uhlenbeck process can be adapted to generate a long memory feature if there is superposition of an infinite number of these processes creating a long memory diffusion.

As the series' are fractionally integrated, the implied volatility from the one-day estimates at different aggregate levels using the variance of partial sums property can be estimated. Specifically using $|r_t|$ as an example, this states that a fractionally integrated variable follows a scaling law of the form $Var(|r_t|) = cT2^{d+1}$. This allows us to infer the implied variance at different aggregations levels, T, assuming knowledge of the unconditional variance at one level. To investigate whether this is correct, the logarithm of the variance of the partial sum of the daily absolute return realisations, $ln(Var(|r_t|_T)$, are regressed against the logarithm of T at different aggregation levels, T = 1, 2, …, 40. The slope of the regression, $d_A$, are given in table III with the corresponding standard errors. Findings for $d_A$ and $d_{GPH}$ are similar. Also, in figure 5, the linear fit for the regression between the logarithm of the variance of the partial sum of the daily absolute return realisations against the logarithm of T is presented. The goodness of fit is clear with the R-squared for each asset's relationship being in excess of 0.990. This association between one-day volatility estimates and their implied T aggregation estimates in the context of the risk management estimates is discussed in the next section.

INSERT FIGURE 5 HERE

**V. Aggregate Measures Applied to Minimum Capital Requirements:**

Having first introduced daily volatility and returns measures they are now used in a risk management application. To illustrate, rescaled returns, $[z_t]$, are obtained by dividing the daily returns by the daily standard deviations, $[r_t]/[\sqrt{r_t^2}]$. The series captures the pattern of dependence in the futures contracts. As indicated in table I, each of the futures contracts returns exhibit fat tails which would lead to an underestimation of risk quantiles under gaussian assumptions. This is because too many realisations lie outside different quantile grids, for example 95%, relative to the bell-shaped normal distribution. Moving to the distributional properties of the rescaled returns excess skewness is removed for all contracts. More importantly, the fat tail feature has been fully removed from the original returns as the negative coefficient indicates a thin tailed distribution as can be clearly seen in figure 6. Particularly for the bond contract, there are now gaussian features with insignificant skewness and kurtosis. Also for the other contracts, assuming normality will result in conservative risk management estimates suitable for risk averse investors as the percentiles of the rescaled returns belong to a relatively thin tailed distribution vis-à-vis normality.[19] Now, excess observations lie inside different quantile grids in comparison to the normal distribution.

INSERT FIGURE 6 HERE

Turning to the risk management application, the paper finds that the minimum capital requirement for investment purposes varies according to different classifications of assets. As futures are highly leveraged assets, investors should face prudential controls from regulators on the losses that can incur. Protection against losses would also be affected by the degree of investor' risk averseness. The actual minimum capital deposit is made up of short-term liquid assets holdings and prearranged credit available to the investor. The use of any capital deposit depends on the conditional trading environment as opposed to the unconditional one as futures are marked to market daily, and individual daily losses rather than average losses over the lifetime of the investment will determine whether the capital deposit is adequate or not. Actual minimum capital requirements are measured according to the fluctuations in the price series'. Large fluctuations require greater adjustments for the deposits given by investors.

Turning specifically to the risk management application, and specifically investigating the probability of the minimum capital deposit being sufficient to cover a large proportion of all possible price movements that the asset may experience. These price movements may occur for (net) long and (net) short positions. This variable is denoted as the loss-covered probability. It is a VaR type exposition examining potential losses resulting from a realized volatility estimate at a certain probability. First looking at the short position, suppose an investor has available a capital deposit of $Sr_{mincapital}$ expressed as a percentage of their total investment, and they want to determine if this will cover a large proportion, for example 99%, of all losses $Sr_{loss}$ again expressed as a percentage:[20]

(11) $$P[Sr_{loss} < Sr_{min capital}] = 0.99$$

This implies that the investor would have a capital deposit that covers 99% of all price movements on their short position. Using (11), devise a one-day forecast of the capital required as a percentage of the total investment for the short position by using the realized volatility measures and the rescaled returns series. Let $\lambda_t$ equal the minimum capital percentage requirement on a short position, and this is related to the realized volatility measure as follows: [21]

(12) $$\lambda_t = \exp(\sqrt{r^2_{t+1}} z_q) - 1$$

with a one period forecast of the realized volatility and the quantile (99%) of the rescaled returns series. Any quantile $z_q$ for a short position would be on the upside of the distribution of realisations and would have a positive value that is in contrast to negative ones for a long position on the downside of the distribution. $[\sqrt{r_{t+1}^2}]$ accounts for more recent volatility levels using an average of the window of 20 lags upto t.

Investors holding a short position hold negative quantities of the asset, and any price increase represents a loss. Price fluctuations in excess of the loss-covered probability are denoted as the loss-exceeded probability, and these should occur with a 1% frequency:

(13) $$P[Sr_{loss} > Sr_{mincapital}] = 0.01$$

Turning to a long position and express the minimum capital requirement $Lr_{mincapital}$ as a percentage of their total investment. Now, a long position implies the investor holding a positive quantity of the asset. Assuming the investor wants to know whether the capital deposit will cover all possible losses $Lr_{loss}$ at a certain probability:

$$P[L_{loss} < Lr_{min\,capital}] = 0.99 \qquad (14)$$

Again, express the minimum capital requirement as a percentage of the total investment for the long position by using the realized volatility measures and rescaled returns series. Let $v_t$ equal the minimum capital percentage requirement on a long position and this is related to the realized volatility measure as follows:

$$n_t = 1 - \exp(\sqrt{r_{t+1}^2} z_q) \qquad (15)$$

Also, price fluctuations in excess of the loss-covered probability are denoted as the loss-exceeded probability, and these should occur with a 1% frequency:

$$P[Lr_{loss} > Lr_{min\,capital}] = 0.01 \qquad (16)$$

The actual amount for the minimum capital requirement will depend on the value of the futures contracts invested in (number of contracts * price of contracts) and the actual price change in the futures. A priori it is expected that the capital deposit is greater for a short position than a long one as the limiting distribution of futures prices are bounded on the downside of the distribution by zero, whereas there is no upper bound on the upside distribution.

Dealing specifically with the rescaled returns in table IV the other parameter required $\sqrt{r_{t+1}^2}$ is measured as 1.009% (FTSE100), 0.468% (UK Long Gilt), and 0.022% (Sterling). The minimum capital requirement $Sr_{mincapital}$ expressed as a percentage of the total investment can be given at any confidence level. For example, to cover 90% of all price fluctuations on a short position, the investor would require a deposit of 1.397%, 0.529% and 0.024% for the stock index, bond and interest rate contracts respectively. As it is assumed that the sampling interval is near continuous, the measurement error is negligible. This loss-covered probability is dependent on the inherent volatility levels of each asset with riskier assets resulting in higher values and consequently higher deposits. Also, moving to a higher safety level would result in higher capital requirements. Turning to a long position, slightly smaller capital requirements occur for the loss-covered probability of 90% with values of 1.193%, 0.486% and 0.022% for the FTSE100, UK Long Gilt, and Sterling futures respectively. It is also beneficial to use the loss-exceeded probability, which suggests for example in the case of the FTSE100, that a minimum capital requirement of 1.193% as a percentage of total outlay would be insufficient for 10% of all outcomes. As can be seen in table IV, the minimum capital requirements increase at higher confidence levels. Note also that the long and short position requirements are similar in magnitude reflecting the insignificant skewness in the rescaled returns series. Gaussian estimates are also presented in table IV for comparison purposes and again the magnitude of values are similar to the long and short position values reflecting the removal of the fat-tailed property from the assets' returns series' and their almost gaussian behaviour.

INSERT TABLE IV HERE

To date the discussion has concentrated on an individual day context. Often risk management decisions and their related estimates are necessary for longer horizons, T, and a framework is now provided that will complete this task accurately.[22] Alternative methods are available for obtaining long horizon estimates from shorter ones including scaling in the case of normality and temporal aggregation rules in the case of GARCH processes (see Drost and Nijman, 1993). Due to the long memory feature in the returns series, the fractionally integrated scaling law rule is used to estimate the implied volatility at different horizons, T.

As seen in figure 5, the assets analysed follow the variance of partial sums property by offering very good fits between the logarithm of aggregation levels and the logarithm of the variance of the partial sum of the absolute realisations. Hence scaling up by $cT2^{d+1}$ using the unconditional variance for one-day aggregate measures gives accurate volatility measures for 40 days in this study. To generalize, by following the rescaling procedure outlined earlier offers correctly calculated associated minimum capital requirements.

**VI. Summary and Conclusion:**

This paper examines model-free volatility estimates for three different asset types using UK futures. These estimates are then incorporated into minimum capital requirements for both long and short positions. Regulators face a dilemma in setting requirements between minimising the losses due to security firms not fulfilling their contracts, and maximising the opportunities for a competitive environment, encouraging investor participation. Whilst alternative approaches are used in practice to set capital deposit levels, this paper adopts the principals of portfolio theory in a VaR type model that measures the risk exposure facing investors with statistical measures of volatility. Various minimum capital requirement estimates are provided that measure the potential of default at various probability levels.

The key to imposing appropriate minimum capital requirements on a daily basis requires accurate estimates of realized volatility for different assets. Using these estimates as inputs, the regulator can balance the trade-off between offsetting security firm default and encouraging financial trade in an optimal manner. Using high frequency realisations, this paper provides alternative, but accurate, estimates of volatility related to statistical implementation of standard portfolio theory. The measures are based on discrete estimation of aggregated high frequency realisations underpinned by a continuous time framework that varies the measurement error according to the interval size used. In contrast, the standard application is to assume variables belong to a known probability distribution function or to estimate volatility using discrete time series models with the associated statistical bias. The choice of very high frequency realisations using 5-minute intervals minimises the bias and can be treated as error free. Squared returns provide model-free estimates of volatility relying on the quadratic variation of a diffusion process that equates this latent variable with the squared realisations. In addition, absolute returns are incorporated into the methodology due to their robustness in the presence of fat tails, and given the motivation to calculate risk management estimates based on the probability distribution of outcomes.

The FTSE100, the UK Long Gilt and the Sterling futures' are the most actively traded and thus are chosen as the UK representatives of stock index, bond and interest rate asset types. Daily realized volatility estimates are obtained using cumulative intraday realisations and a number of interesting findings are reported. First unconditionally, daily realized volatility estimates like returns indicate non-gaussian features with excess skewness and kurtosis. Second conditionally, the realized volatility estimates follow previous evidence by varying over time with clustered period of high and low values.

Third and more importantly, the realized volatility estimates incorporate the long memory property showing a dependency structure that has a specific decay pattern in line with a fractionally integrated series. Within this dependency structure, a cyclical pattern occurs corresponding to a self-similar shape across individual days. Using these alternative risk estimates, the dynamics of the volatility variable are adequately captured and the paper obtains rescaled returns that are near gaussian. Fourth, these rescaled returns are applied to calculate minimum capital requirements for the futures analysed. Portfolios containing stock index assets would incur more risk, and thus require a larger capital deposit, than the other two assets analysed. Moreover, the previously overlooked interest rate asset type is the least volatile. Given the time series properties of the daily volatility series' a simple extension to calculate accurate volatility forecasts over longer investment horizons that follow a scaling law for fractionally integrated series is suggested.

Table I: Summary Statistics for Daily Returns, [$r_t$], of UK Futures Series

|  | FTSE | UK Long Gilt | Sterling |
|---|---|---|---|
| Min | -5.484 | -2.686 | -0.484 |
| 1st Qtr | -0.593 | -0.166 | -0.032 |
| Median | 0.103 | 0.009 | 0.000E+00 |
| 3rd Qtr | 0.721 | 0.202 | 0.032 |
| Max | 9.981 | 1.784 | 1.235 |
| Mean | 0.071 | 0.010 | -1.424E-04 |
| Variance | 0.015 | 0.002 | 5.459E-05 |
| Std. Deviation | 1.233 | 0.484 | 0.007 |
| Skewness | 0.319* | 0.225* | 0.775* |
| Kurtosis | 5.668* | 1.460* | 8.825* |
|  |  |  |  |
| Percentiles |  |  |  |
| 0.5 | -3.657 | -1.126 | -0.301 |
| 1.0 | -2.918 | -1.034 | -0.247 |
| 5.0 | -1.947 | -0.832 | -0.108 |
| 10.0 | -1.337 | -0.713 | -0.065 |
| 90.0 | 1.413 | 0.716 | 0.066 |
| 95.0 | 1.970 | 0.814 | 0.096 |
| 99.0 | 3.214 | 0.977 | 0.192 |
| 99.5 | 3.833 | 1.280 | 0.226 |

Notes: The daily return measure, [$r_t$], is outlined in the main text. With the exception of skewness and kurtosis coefficients, all values are expressed in percentage form. Normal iid skewness and kurtosis values should have means equal to 0, and variances equal to 6/T and 24/T respectively. Thus, standard errors for the skewness and kurtosis parameters are 0.077 and 0.154 respectively. Significant kurtosis indicates excess kurtosis vis-à-vis normality. The symbol * represents significant skewness and kurtosis using two standard errors.

Table II: Summary Statistics for Daily Realized Volatility Measures, using Aggregated Returns of UK Futures Series

|  | FTSE | UK Long Gilt | Sterling | FTSE | UK Long Gilt | Sterling |
|---|---|---|---|---|---|---|
| $[r_t^2]$ |  |  |  | $Ln[r_t^2]$ |  |  |
| Mean | 0.014 | 0.009 | 0.000 | -9.268 | -10.857 | -14.809 |
| Variance | 0.000 | 0.000 | 0.000 | 0.765 | 3.905 | 0.871 |
| Std. Deviation | 0.018 | 0.014 | 0.000 | 0.874 | 1.976 | 0.933 |
| Skewness | 6.553* | 2.416* | 8.436* | 0.253* | 0.090 | 0.279* |
| Kurtosis | 73.218* | 6.364* | 88.240* | -0.121 | -0.991* | 2.648* |
|  |  |  |  |  |  |  |
| $[\sqrt{r_t^2}]$ |  |  |  | $Ln[\sqrt{r_t^2}]$ |  |  |
| Mean | 1.071 | 0.683 | 0.066 | -4.634 | -5.429 | -7.405 |
| Variance | 0.003 | 0.004 | 0.000 | 0.191 | 0.976 | 0.218 |
| Std. Deviations | 0.522 | 0.628 | 0.044 | 0.437 | 0.988 | 0.467 |
| Skewness | 2.014* | 1.292* | 3.520* | 0.253* | 0.090 | 0.279* |
| Kurtosis | 8.506* | 0.865* | 20.610* | -0.121 | -0.991* | 2.648* |
|  |  |  |  |  |  |  |
| $|r_t|$ |  |  |  | $Ln|r_t|$ |  |  |
| Mean | 7.339 | 2.994 | 0.265 | -2.704 | -3.780 | -6.109 |
| Variance | 0.115 | 0.049 | 0.000 | 0.184 | 0.605 | 0.651 |
| Std. Deviation | 3.389 | 2.211 | 0.181 | 0.428 | 0.778 | 0.807 |
| Skewness | 1.776* | 1.652* | 1.498* | 0.097 | 0.382* | 2.883* |
| Kurtosis | 7.254* | 4.167* | 4.955* | -0.061 | -0.046 | 52.588* |
|  |  |  |  |  |  |  |
| $|r_t^{0.5}|$ |  |  |  | $Ln|r_t^{0.5}|$ |  |  |
| Mean | 234.912 | 123.930 | 22.995 | 0.820 | 0.130 | -1.662 |
| Variance | 37.543 | 22.659 | 2.297 | 0.074 | 0.191 | 0.528 |
| Std. Deviation | 61.272 | 47.602 | 15.155 | 0.272 | 0.437 | 0.727 |
| Skewness | 0.426* | 0.221* | 0.959* | 0.788* | 1.191* | 0.837* |
| Kurtosis | 0.862* | -0.861* | 1.869* | 3.679* | 6.475* | 0.935* |

Notes: The daily realized volatility measures are outlined in the main text. All values for $[r_t^2]$, $[\sqrt{r_t^2}]$, $|r_t|$, and $|r_t^{0.5}|$, are expressed in percentage form with the exception of the skewness and kurtosis coefficients. Normal iid skewness and kurtosis values should have means equal to 0, and variances equal to 6/T and 24/T respectively. Thus, standard errors for the skewness and kurtosis parameters are 0.077 and 0.154 respectively. Significant kurtosis indicates excess kurtosis vis-à-vis normality. The symbol * represents significant skewness and kurtosis using two standard errors.

Table III: Memory in the Daily Realized Volatility Measure, $|r_t|$, using Aggregated Returns of UK Futures Series

|  | FTSE | UK Long Gilt | Sterling |
|---|---|---|---|
| Ljung-Box (40) | 15525 | 13304 | 7769 |
| $d_{GPH}$ | 0.443 | 0.406 | 0.381 |
| $d_A$ | 0.471 | 0.439 | 0.416 |
|  | (0.002) | (0.006) | (0.005) |

Notes: All Ljung-Box statistics are significant. The standard errors for the $d_{GPH}$ are 0.040 as outlined in the text. The standard errors for $d_A$ are presented in parentheses. All $d_{GPH}$ estimates are significantly different from zero thereby rejecting a short memory characteristic.

Table IV: Daily Minimum Capital Estimates for Short and Long Positions in UK Futures

| Probability | FTSE | UK Long Gilt | Sterling |
|---|---|---|---|
| Long Position | | | |
| 90% | 1.194 | 0.486 | 0.022 |
| 95% | 1.502 | 0.610 | 0.028 |
| 99% | 2.033 | 0.888 | 0.040 |
| 99.5% | 2.202 | 0.934 | 0.043 |
| | | | |
| Short Position | | | |
| 90% | 1.397 | 0.529 | 0.024 |
| 95% | 1.718 | 0.631 | 0.030 |
| 99% | 2.489 | 0.831 | 0.039 |
| 99.5% | 2.683 | 0.968 | 0.041 |
| | | | |
| Normal | | | |
| 90% | 1.293 | 0.600 | 0.028 |
| 95% | 1.660 | 0.770 | 0.036 |
| 99% | 2.347 | 1.089 | 0.051 |
| 99.5% | 2.599 | 1.205 | 0.057 |

Notes: The minimum capital requirements are expressed as a percentage of the total investment. Results are presented individually for the long and short positions using the daily realized volatility estimates as discussed in the text. Estimates are presented assuming normality for comparison purposes where long and short position values are equivalent.

Figure 1: Box Plots for UK Futures Realized Volatility Series.

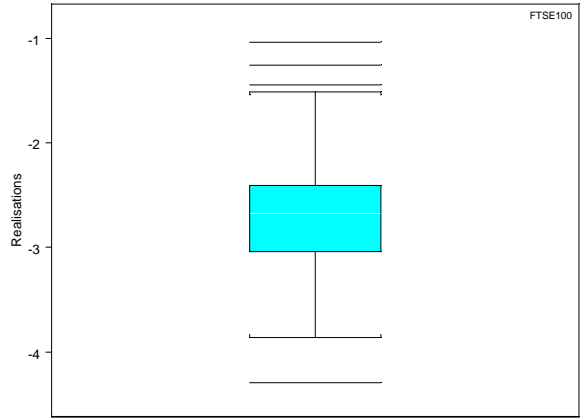

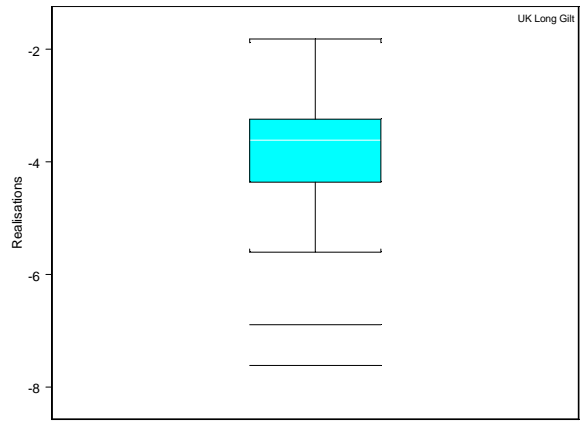

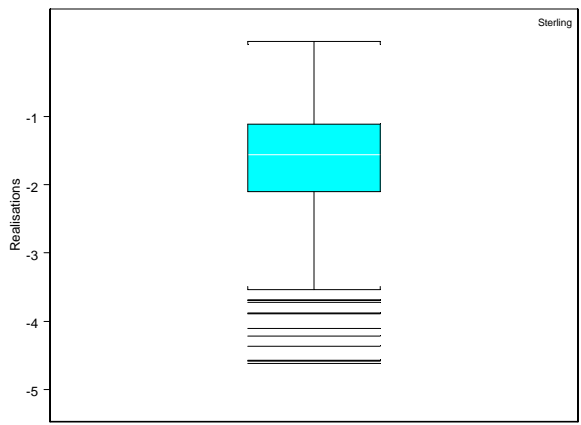

Notes: The realized volatility series chosen for presentation are based on those with the optimal skewness and kurtosis coefficients vis-à-vis normality. Specifically, the $Ln|r_t|$ series for the FTSE100 and UK Long Gilt contracts and the $Ln|r_t|^{0.5}$ series for Sterling are chosen. The upper and lower quartile grids are given by the line segments with the spikes at either end.

Figure 2: Time Series Plots of UK Daily Futures Realized Volatility Series, |r$_t$|

FTSE100

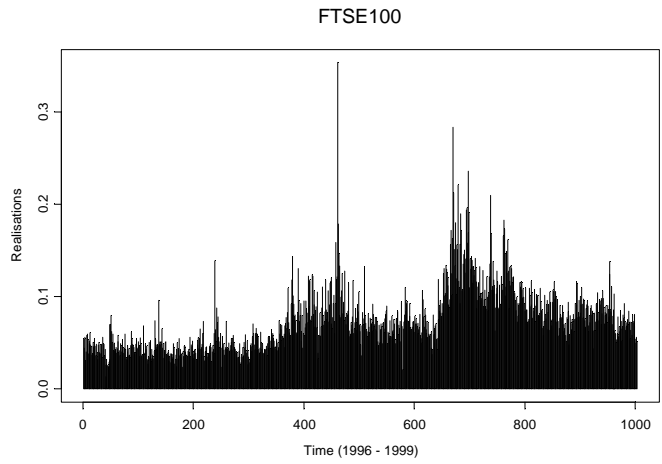

UK Long Gilt

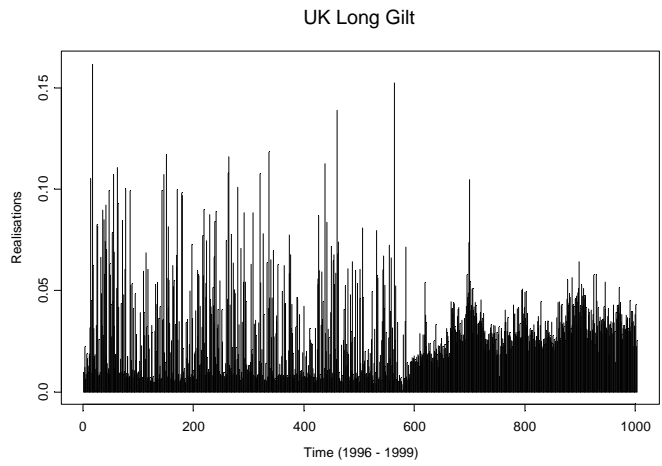

Sterling

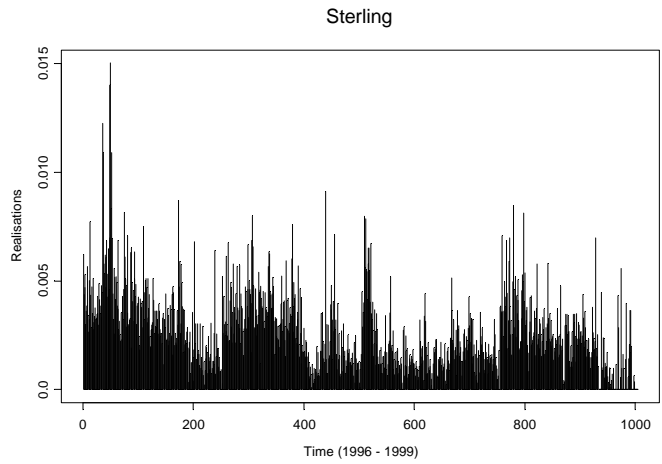

Figure 3: Plots of Autocorrelation Values for UK Futures Daily Realized Volatility Series, |$r_t$|, for 50 Daily Lags.

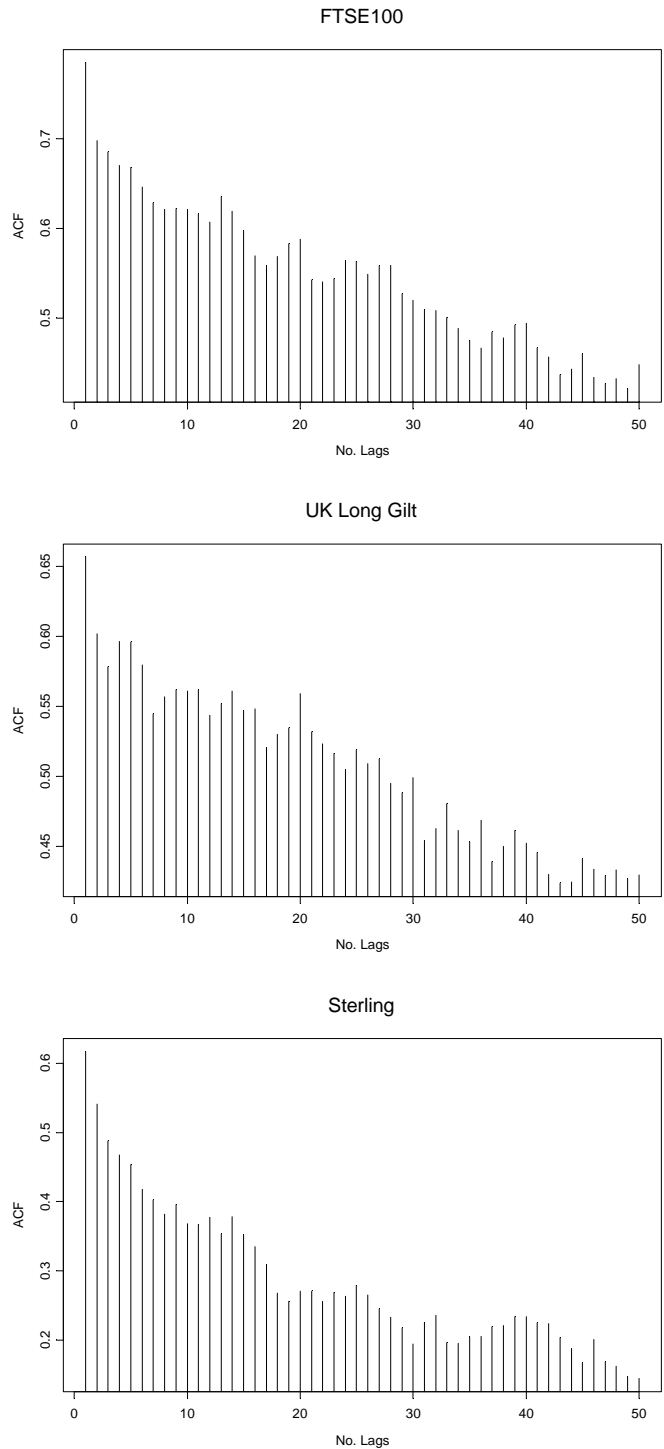

Notes: Each asset contains a different number of daily intervals -FTSE100 (113), UK Long Gilt (120), and Sterling (118). The critical value for each contract is 0.062 as estimated by $\pm 1.96/\sqrt{T}$.

Figure 4: Plots of Autocorrelation Values for UK Futures Daily Realized Volatility Series, |$r_t$| across 5 Days.

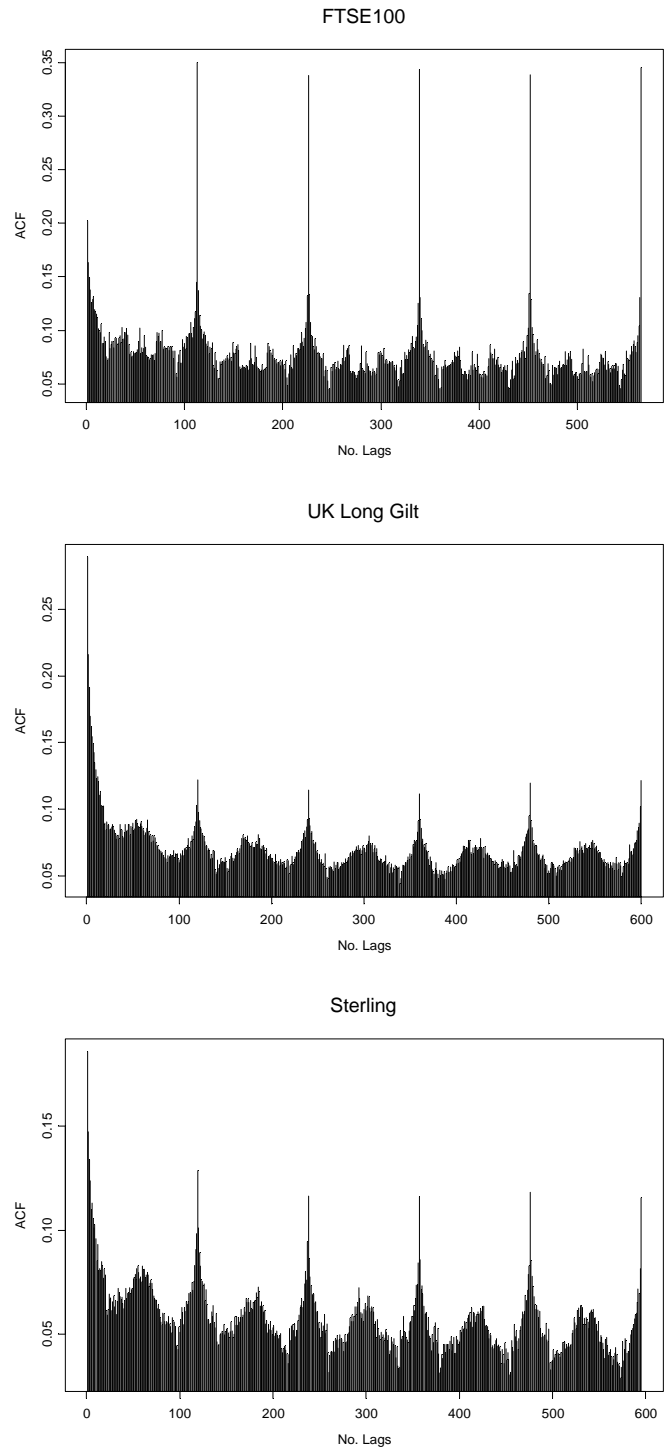

Notes: Each asset contains a different number of daily intervals -FTSE100 (113), UK Long Gilt (120), and Sterling (118). The critical value for each contract is 0.062 as estimated by ±1.96/√T.

Figure 5: Linear Fit Plots of implied aggregated realized volatility using UK Futures Daily Volatility Series, $|r_t|$, against different aggregation levels.

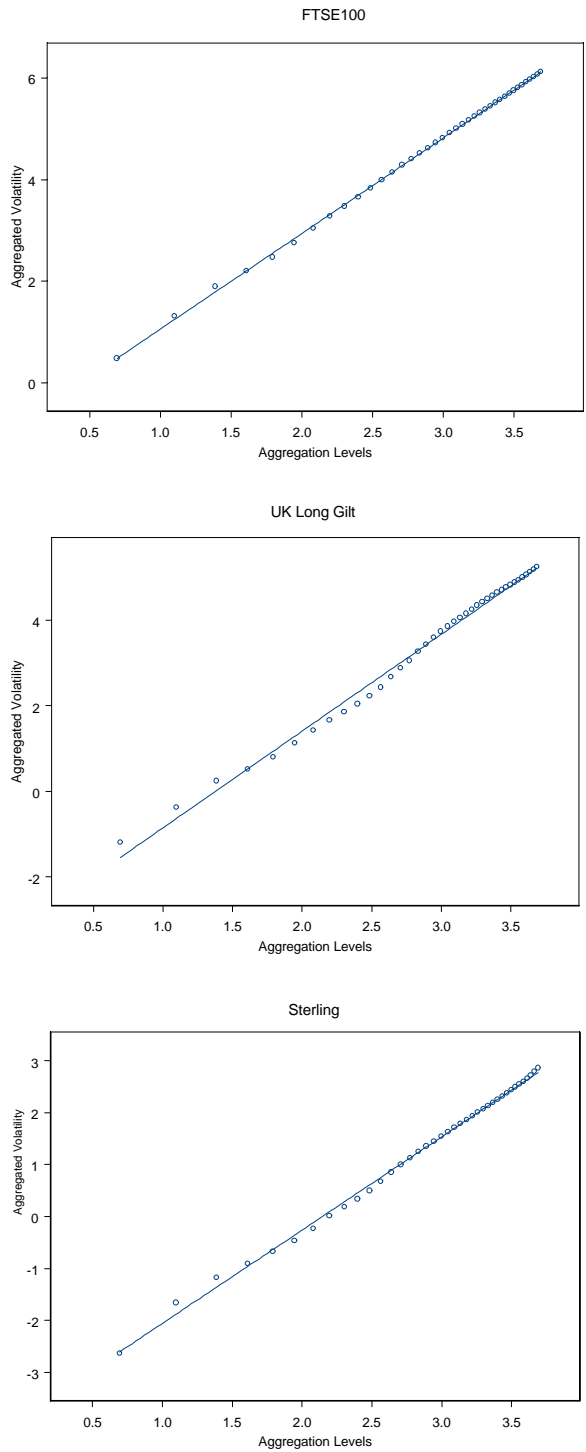

Notes: Aggregated volatility represents the logarithm of the variance of the partial sum of the daily absolute return realisations, $\ln(\mathrm{Var}(|r_t|_T))$, and aggregation levels the logarithm of T at different aggregation levels, T = 1, 2, …, 40.

Figure 6: Box Plots for UK Futures Daily Rescaled Returns Series, $[z_t] = [r_t]/[r\sqrt{}_t^2]$

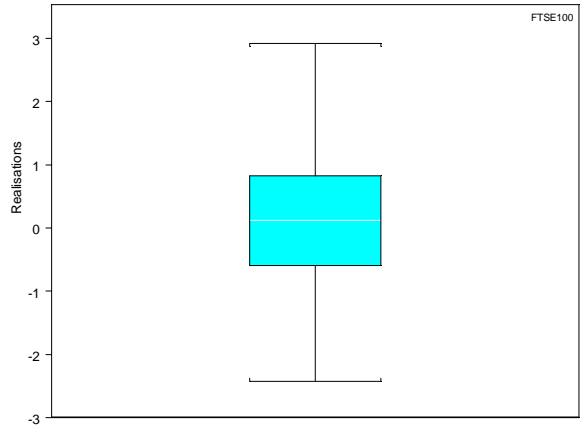

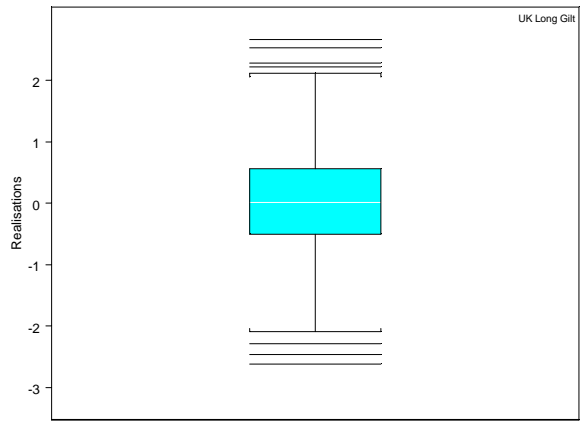

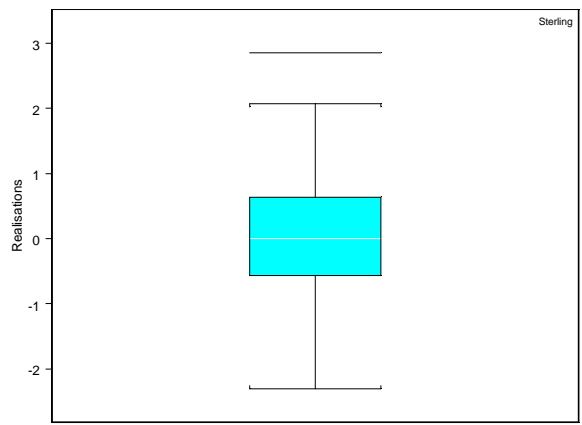

Notes: the line segments give the upper and lower quartile grids with the spikes at either end.

[1.] Previously Areal and Taylor (2002) present a single realized volatility measure for squared realizations.

[2] Dimson and Marsh (1997) provide a comprehensive discussion of the procedures in place for measuring minimum capital requirements.

[3] For a comprehensive discussion of comparative methodologies for measuring minimum capital requirements see Hsieh (1993).

[4] This study is currently being extended to examine minimum capital requirements for financial firms.

[5] For example, the UK's Securities and Investment Board (1987) suggests, "The primary objective … is to ensure that the risks which a firm undertakes are not disproportionate to its resources. It is *not* that there would be complete confidence that a market participant will *never* default, but that the size and frequency of any failures should not have material systemic consequences and the risk of loss to retail investors… should be small".

[6] In practice, measurement error will occur as high frequency discrete realizations are substituted into the continuous time process, but the error size is controlled through varying the frequency of realizations used. In the risk management estimates reported, the findings of simulations by Andersen and Bollerslev (1998) are incorporated by assuming that using a 5-minute interval results in a miniscule measurement error.

[7] Hence $r_{1,t} = p_t - p_{t-1/1}$, which can easily be scaled for different holding periods.

[8] Merton (1980) suggests that for $m \to \infty$, precise volatility estimates can be obtained.

[9] The actual holidays fell on slightly different calendar dates each year depending on whether the respective holiday fell on a weekday or not.

[10] The impact from using different interval size is currently being investigated by the author.

[11] In contrast, individual asset returns experience negative correlations.

[12] Alternative procedures could be followed in the choice of contract used for analyzing data and this paper follows a commonly applied technique (Taylor, 1986).

[13] Using values from the previous intervals occurred with a frequency of less than 1% for the FTSE100 and Long Gilt contracts, whereas this was followed by approximately 10% of the time for the Sterling futures. There would be a negligible impact on the aggregation process from this given the large number of observations used.

[14] Sample variances are calculated using $\dfrac{\sum_{t=1}^{T}[r_t - \bar{r}]^2}{T-1}$ which simplifies to $\dfrac{\sum_{t=1}^{T}[r_t^2]}{T-1}$ assuming zero mean, whereas the mean value of the volatility estimate in table II has a denominator of T.

[15] Throughout the paper a selection of findings are presented only. All results are available on request.

[16] For the null hypothesis of gaussian iid variables to hold, the sample skewness and kurtosis values should have means equal to 0, and variances equal to 6/T and 24/T respectively.

[17] All the returns series indicate volatility clustering and closely resemble a white noise series. These features are common for financial time series and have been noted for the series' analysed using observations gathered at a daily interval. Results are available on request.

[18] For example, based on the presented estimates in table III of the $|R_t|$ series' for $d_{GPH}$ of 0.443 (FTSE100) results in $\delta < 0.639$, and 0.381 (UK Long Gilt) results in $\delta < 0.604$.

[19] Leptokurtosis disappears as the FTSE100 and Sterling contracts have kurtosis coefficients of –0.411 and –0.428 respectively.

[20] Any quantile could have been chosen with a more (less) risk averse investor choosing a higher (lower) probability of safety, and 99% is for illustrative purposes only.

[21] The volatility measure $[\sqrt{r_{t+1}^2}]$ is used as an example.

[22] Whilst it is not the motivation of this paper to determine of the optimal holding period for risk management decision making, examples of long horizons used include the minimum 10-day holding period suggested by the Basle's Committee 1996 'ammendment to the capital accord to incorporate market risks' and the one year Bankers Trust RAROC system (see Diebold et al (1988)).